\documentclass[11pt,a4paper]{article}
\usepackage[T1]{fontenc}
\usepackage{lmodern}
\usepackage[margin=25mm]{geometry}
\usepackage{amsmath,amssymb,bm,graphicx,booktabs,array}
\usepackage[numbers,sort&compress]{natbib}
\usepackage{microtype}
\usepackage[font=small,labelfont=bf]{caption}
\usepackage{placeins}
\usepackage[colorlinks=true,linkcolor=blue,citecolor=blue,urlcolor=blue]{hyperref}
\graphicspath{{figures/}}
\newcommand{\pw}{{p_w}}
\newcommand{\vs}{{\bm v_s}}
\newcommand{\sig}{\bm\sigma}
\newcommand{\sige}{\bm\sigma'}
\newcommand{\Id}{\bm I}
\title{A $u$--$p_w$ SPH framework with mechanical--hydraulic boundary treatment in GeoDualSPHysics for large-deformation geomechanics}
\author{\normalsize Yangfan Ma\thanks{Department of Civil and Structural Engineering, Kyushu University, Fukuoka, Japan.}\quad
Bin Su\thanks{School of Civil Engineering, Central South University, Changsha, 410075, China.}\quad
Zheng Han\footnotemark[2]\quad
Changze Li\thanks{Graduate School of Science, Kyoto University, Kyoto, Japan.}\quad
Mitsuteru Asai\footnotemark[1]}
\hypersetup{pdfauthor={Yangfan Ma, Bin Su, Zheng Han, Changze Li, Mitsuteru Asai},
pdftitle={A u-pw SPH framework with mechanical-hydraulic boundary treatment in GeoDualSPHysics for large-deformation geomechanics}}
\date{}

\begin{document}
\maketitle
\begin{abstract}
Pore-water pressure changes the effective stress and deformation of saturated geomaterials, while skeleton deformation generates pressure and drives drainage. Resolving this feedback together with hydraulic boundary conditions is necessary for simulations that connect consolidation to large post-failure movement. This paper presents a hydromechanical extension of GeoDualSPHysics using a single-layer displacement--pore-water-pressure ($u$--$p_w$) smoothed particle hydrodynamics (SPH) framework. An explicit pressure-rate equation combines volumetric compression and Darcy diffusion, and the resulting pore pressure feeds back into the skeleton momentum balance. The implementation extends the existing modified dynamic boundary condition (mDBC) infrastructure to carry hydraulic ghost values alongside mechanical boundary data. Prescribed pressure at drained surfaces is combined with impermeable walls and impermeable loaded contacts, with separate control of loading and drainage activation. The formulation is assessed through one-dimensional consolidation, two-dimensional flexible-strip loading, and the three-dimensional Cryer problem. The Cryer simulations reproduce the non-monotonic central pressure response over four Poisson ratios, with normalized pressure root-mean-square errors of 0.012--0.030. The consolidation tests also expose remaining settlement and late-time pressure discrepancies. Applications to 5 m and 8 m sensitive-clay slopes illustrate sequential retrogressive failure and large deformation. The contribution is the integration of hydraulic feedback and boundary handling within GeoDualSPHysics; the results establish an initial verification basis while identifying remaining accuracy and convergence questions.
\end{abstract}

\section{Introduction}
Saturated geomaterials can pass from a nearly equilibrated state through localized failure to large displacement and flow. Throughout this transition, changes in pore-water pressure modify effective stresses and hence the forces carried by the soil skeleton. Conversely, skeleton compression and extension generate pressure changes, and hydraulic gradients redistribute water. A computational description of this feedback must also distinguish between drained surfaces, impermeable supports, and loaded contacts. These requirements are especially relevant to consolidation beneath foundations and the progressive destabilization of sensitive-clay slopes.

Large-deformation numerical methods provide several routes to this problem. Particle finite element formulations combine moving discretizations with coupled soil--water equations, as demonstrated by \citet{jin2022}. SPH follows material motion without a fixed connectivity and is therefore attractive for evolving free surfaces, localization, and post-failure movement. Elastoplastic geomechanical SPH established a basis for simulating soil failure \citep{bui2008}; subsequent developments addressed coupled equations through fractional-step schemes \citep{blanc2012} and saturated-soil response with hydrostatic pore pressure \citep{bui2013}. Prescribed hydrostatic pressure, however, does not describe transient pressure generation and dissipation caused by deformation and drainage.

Coupled SPH has developed along two complementary particle-discretization routes. In a \emph{multi-point multiphase} approach, also termed multi-layer SPH, the soil and water phases occupy separate, overlapping particle sets, each with its own kinematic description. Bui and Nguyen's two-layer formulation \citep{bui2017} solves the phase equations on different Lagrangian discretizations and couples them through pressure and interaction forces. Separate phase trajectories provide a direct description of relative motion and can support seepage coupled to external free-water flow. This flexibility entails additional phase variables, neighbour searches, and interpolation between particle sets; adequate cross-phase support must be maintained as the phases move relative to one another. Its computational cost and suitability depend on the phase equations and interaction treatment used.

In a \emph{single-point multiphase} approach, or single-layer SPH, each material point carries the soil and pore-fluid fields over the same representative mixture volume, and one particle set follows the skeleton. Relative water transport is represented by additional fluid variables or flux equations, so co-location does not impose zero seepage. This route avoids transporting a separate pore-water particle set and provides a common moving geometry for mechanical and hydraulic boundary treatment. Its development includes strong-coupling ISPH \citep{morikawa2022}, single-layer descriptions for unsaturated slope failure \citep{feng2022coupled}, stabilized saturated $u$--$p_l$ formulations \citep{lian2023}, and Darcy-based pore-pressure evolution applied to sinkhole development \citep{chen2024}. General single-layer SPH work has also addressed coupled seepage operators, stabilization, and hydromechanical boundary conditions for saturated and unsaturated porous media \citep{feng2024coupled}. These developments are complementary to multi-layer methods. Co-located formulations require consistent treatment of phase transport and boundary data, and the ability to represent independent free-water motion or large relative fluid inertia depends on the retained equations and any additional phase representation.

Particle layout, physical approximation, and pressure solution strategy must therefore be distinguished. Reduced $u$--$p$ descriptions use skeleton displacement and pore pressure as primary fields, whereas $u$--$w$--$p$ and $u$--$U$--$p$ descriptions retain additional relative-flow or fluid-displacement variables. The meaning of $w$ and $U$ follows each formulation and should not be inferred from notation alone. A single particle set can carry additional fluid unknowns; the number of particle sets alone does not determine whether fluid inertia is retained. Morikawa and Asai's strong-coupling ISPH method uses a semi-implicit projection and an embedded pore-fluid description \citep{morikawa2022}. Lian et al. developed a stabilized $u$--$p_l$ framework and a time-stepping strategy to address the restrictive flow-related increment \citep{lian2023}. Yao et al. compared $u$--$w$--$p$ and $u$--$p$ SPH formulations and investigated boundary conditions and volumetric locking \citep{yao2025}. Their distinction between moderate and strongly dynamic loading also highlights the limits of reduced formulations. More general $u$--$U$--$p$ methods explicitly address the fluid momentum and drag-force discretization \citep{wang2026}. Recent work further emphasizes consistency and stabilization in coupled SPH \citep{tong2026}.

The single-point route is adopted here because the intended problems couple large skeleton deformation to pore-pressure generation and Darcy drainage within saturated soil, without separately tracking an external water body. Placing pressure and skeleton variables on GeoDualSPHysics particles allows the extension to reuse its stress update, neighbour structure, surface tracking, and mDBC geometry. It also permits pressure and no-flux conditions to be coordinated with mechanical slip on the same evolving boundary representation. The present reduction neglects a separate relative-fluid acceleration equation; this is a physical approximation in addition to the particle-layout choice. Likewise, saturation is an assumption of this implementation, rather than an inherent restriction of all single-point multiphase methods. The resulting implementation rationale is to retain the required feedback and boundary distinctions with fewer particle sets; quantitative performance gains still require a matched comparison.

The present pressure-rate formulation is closely related to the explicit PR formulation of \citet{castillo2026}, who introduced both pressure-Poisson and pressure-rate single-layer SPH schemes and examined consolidation, Cryer's problem, and retrogressive failure. Building on this formulation, the present work integrates hydraulic feedback into GeoDualSPHysics while retaining its established geomechanical discretization and boundary infrastructure.

GeoDualSPHysics provides a GPU-oriented SPH implementation for large-deformation geomaterials, including stress treatment and mechanical mDBC support \citep{feng2026geodual}. The original mDBC construction uses interior ghost locations to recover boundary fields \citep{english2022}. In a saturated-soil extension, recovering mechanical stress and velocity is insufficient: the Darcy operator also needs hydraulic values consistent with the intended pressure or flux condition. A wall may be mechanically free-slip yet hydraulically impermeable, while the loaded part of a free surface may remain impermeable after the adjacent surface begins to drain. Treating these choices independently within one boundary representation is the principal implementation focus of this work.

This paper therefore presents three connected developments: an explicit $u$--$p_w$ extension of GeoDualSPHysics; hydraulic ghost-pressure reconstruction and drained-surface treatment coordinated with mDBC mechanics; and an assessment spanning one-, two-, and three-dimensional coupled tests and retrogressive slope failure. Section~\ref{sec:method} describes the equations and implementation. Section~\ref{sec:tests} presents the consolidation and Cryer tests. Section~\ref{sec:slopes} examines the slope applications. The discussion examines the capabilities demonstrated by the results and the limits of the present verification.

\section{Methodology}\label{sec:method}
\subsection{Saturated-mixture description and hydraulic feedback}
The soil skeleton and pore water occupy a fully saturated continuum. A single particle layer follows the skeleton motion, with position $\bm x$, displacement $\bm u$, velocity $\vs=\dot{\bm u}$, effective stress $\sige$, and pore-water pressure $\pw$. Tension is positive for stress, whereas positive $\pw$ denotes compression. With the incompressible-grain approximation underlying the effective-stress coefficient of unity,
\begin{equation}
 \sig=\sige-\pw\Id,\qquad
 \dot{\bm x}=\vs,\qquad
 \rho\dot{\vs}=\nabla\!\cdot\sige-\nabla\pw+\rho\bm b .
 \label{eq:momentum}
\end{equation}
Here $\rho$ is the mixture density and $\bm b$ is the body acceleration. The mixture momentum retains skeleton motion and inertia, while relative fluid flow is approximated by Darcy's law. A separate fluid-acceleration equation is not solved. The model consequently targets consolidation and low-relative-inertia hydromechanical response; it is not a replacement for a general two-momentum dynamic formulation \citep{yao2025,wang2026}.

Let $k$ denote hydraulic conductivity in m/s, $n$ the prescribed porosity parameter, $K_w$ the pore-water bulk modulus, and $\rho_w$ the reference water density. With elevation $z$ positive upwards and gravitational magnitude $g$, Darcy discharge is
\begin{equation}
 \bm q=-k\nabla h_w,\qquad h_w=\frac{\pw}{\rho_w g}+z .
 \label{eq:darcy}
\end{equation}
For the zero-gravity tests, the elevation contribution is omitted and a reference value of $g$ is retained solely in the conversion between hydraulic conductivity and pressure mobility. The pressure evolution used by the PR implementation is
\begin{equation}
 \dot{\pw}=\frac{K_w}{n}\left[-\nabla\!\cdot\vs-\nabla\!\cdot\bm q\right]
 =\frac{K_w}{n}\left[-\nabla\!\cdot\vs+
 \nabla\!\cdot\!\left\{k\nabla\!\left(\frac{\pw}{\rho_w g}+z\right)\right\}\right].
 \label{eq:pressure}
\end{equation}
This equation follows the compressible-fluid single-layer approach discussed by \citet{castillo2026}. Skeleton compression, $\nabla\!\cdot\vs<0$, raises pressure. Pressure gradients produce discharge, which changes pressure through mass balance. Equation~\eqref{eq:momentum} returns that pressure to the mechanical problem. The term \emph{coupled} refers to this bidirectional feedback; the time advance is explicit rather than a monolithic nonlinear solve.

Porosity and conductivity are specified material parameters in the calculations reported here. A deformation-dependent permeability law, unsaturated flow, and a separate free-water phase are outside the present formulation. The stored reference pressure $p_{w0}$ permits the excess pressure $p_w^e=\pw-p_{w0}$ to be distinguished from a hydrostatic initial field. In the zero-gravity strip test, $p_{w0}=0$ and the pore pressure is also the excess pressure.

\subsection{SPH operators and effective-stress update}
For particle $i$, neighbour volume is $V_j=m_j/\rho_j$, $W_{ij}$ is the smoothing kernel, and $\bm r_{ij}=\bm x_i-\bm x_j$. Wendland kernels are used throughout the numerical tests. The velocity-divergence and diffusion operators have the forms
\begin{align}
 D_i&=\sum_j V_j(\vs_j-\vs_i)\!\cdot\bm G_{ij},\\
 L_i(f)&=2\sum_j V_j(f_i-f_j)
 \frac{\bm r_{ij}\!\cdot\widehat{\nabla}_iW_{ij}}
 {|\bm r_{ij}|^2+\epsilon_h^2},\label{eq:lap}
\end{align}
where $\epsilon_h$ regularizes the denominator. The implementation supports a locally corrected gradient $\widehat{\nabla}_iW_{ij}$; $\bm G_{ij}$ denotes the gradient selected for the compression term. The pressure-rate contribution is assembled from $-D_i$ and the pressure and elevation diffusion terms. Boundary neighbours contribute through their reconstructed mechanical and hydraulic states. The distinction between the compression derivative and the hydraulic ghost value matters: changing a mechanical slip condition must not silently change a wall's prescribed drainage condition.

The momentum calculation combines the effective-stress contribution inherited from GeoDualSPHysics with a symmetric pore-pressure interaction. In the uncorrected pair form the latter is proportional to
\begin{equation}
 \bm a_i^{p}=-\sum_j m_j\frac{\pw_i+\pw_j}{\rho_i\rho_j}\nabla_iW_{ij}.
\end{equation}
The effective stress follows the existing elastoplastic stress-update infrastructure \citep{feng2026geodual}. In the elastic tests, cohesion is set sufficiently high to suppress yielding, giving the objective isotropic elastic rate relation
\begin{equation}
 \overset{\circ}{\sige}=2G\bm D^{\rm dev}+K_d\,\mathrm{tr}(\bm D)\Id,
 \quad G=\frac{E}{2(1+\nu)},\quad K_d=\frac{E}{3(1-2\nu)},
\end{equation}
with $\bm D=\mathrm{sym}(\nabla\vs)$ and an objective stress rate accounting for material rotation. The slopes instead use a strain-softening Drucker--Prager model with zero friction and dilatancy. Their stage-dependent strength is described in Section~\ref{sec:slopes}.

\subsection{Artificial stress and numerical damping}\label{sec:stabilization}
The physical momentum balance in Equation~\eqref{eq:momentum} is supplemented by numerical terms in the particle calculation. Writing $\bm a_i^{\rm eff}$ for the effective-stress interaction, $\bm a_i^p$ for the pore-pressure interaction, and $\bm a_i^{\rm load}$ for imposed surface-loading contributions where applicable, the acceleration is organized as
\begin{equation}
 \begin{aligned}
 \dot{\vs}_i&=\bm a_i^{\rm eff}+\bm a_i^p+\bm b+\bm a_i^{\rm load}\\
 &\quad+\bm a_i^{\rm AS}+\bm a_i^{\rm AV}+\bm a_i^{\rm d}.
 \end{aligned}
 \label{eq:augmentedacceleration}
\end{equation}
Here AS denotes artificial stress, AV artificial viscosity, and d velocity-proportional damping. Their coefficients are numerical settings, separate from the hydraulic conductivity and the constitutive strength parameters.

\paragraph{Artificial stress.}
To reduce tensile instability in cohesive soil, the artificial-stress construction of \citet{bui2008} is applied to the tensile principal components of the skeleton effective stress. Let $\sige_i=\bm Q_i\,\mathrm{diag}(s_{i1},s_{i2},s_{i3})\bm Q_i^{\mathrm T}$, where the columns of $\bm Q_i$ are orthonormal principal directions. With the tension-positive convention, define
\begin{equation}
 \bm R_i=-\frac{\epsilon_{\rm AS}}{\rho_i^2}\,
 \bm Q_i\,\mathrm{diag}\!\left(\langle s_{i1}\rangle_+,
 \langle s_{i2}\rangle_+,\langle s_{i3}\rangle_+\right)\bm Q_i^{\mathrm T},
 \qquad \langle s\rangle_+=\max(s,0).
 \label{eq:artificialtensor}
\end{equation}
Thus compressive principal components do not generate artificial stress. In the uncorrected pair form, its acceleration contribution is
\begin{equation}
 \bm a_i^{\rm AS}=\sum_j m_j f_{ij}(\bm R_i+\bm R_j)\!\cdot\nabla_iW_{ij},
 \qquad f_{ij}=\left[\frac{W(|\bm r_{ij}|,h)}{W(\Delta x,h)}\right]^{q_{\rm AS}}.
 \label{eq:artificialacceleration}
\end{equation}
The dimensionless parameters $\epsilon_{\rm AS}$ and $q_{\rm AS}$ control the correction amplitude and its distance dependence; $q_{\rm AS}$ is distinct from the porosity $n$. The reference distance $\Delta x$ is the initial particle spacing. The kernel weighting strengthens the short-range repulsive contribution as tensile particles approach. At a rigid boundary, $\bm R_j$ is obtained from the reconstructed boundary stress. The correction is added to acceleration and does not alter the constitutive stress update or directly add a term to the pore-pressure equation.

\paragraph{Damping force.}
The velocity-proportional damping of \citet{bui2013}, their Equations (40)--(41), is implemented as
\begin{equation}
 \bm a_i^{\rm d}=-c_{d,i}\vs_i,\qquad
 c_{d,i}=\xi\sqrt{\frac{E}{\rho_i h^2}},\qquad
 \bm F_i^{\rm d}=m_i\bm a_i^{\rm d}.
 \label{eq:damping}
\end{equation}
Here $\xi$ is dimensionless, whereas $c_{d,i}$ has units of s$^{-1}$. The density is the particle mixture density and $h$ is the smoothing length used in that calculation stage. Damping acts on moving soil particles; rigid boundary particles are excluded. Its instantaneous contribution to the kinetic-energy rate is $\dot{\mathcal K}_{\rm d}=-\sum_i m_i c_{d,i}|\vs_i|^2\leq0$. This term removes kinetic energy and assists stress equilibration. In the present elastic benchmarks, the listed damping remains active during the reported transient response; in the slopes, $\xi$ is reduced from 0.02 during prestress to $10^{-5}$ during failure. It is therefore a numerical influence on the transient solution, including pressure generation through the velocity divergence, and is not a calibrated physical soil viscosity or Darcy dissipation. At fixed $\xi$, its rate also scales as $h^{-1}$, which matters when comparing resolutions or changing the smoothing length between stages.

\paragraph{Distinction from artificial viscosity.}
The artificial viscosity uses the linear approaching-particle term of the usual SPH construction \citep{bui2008},
\begin{align}
 \bm a_i^{\rm AV}&=-\sum_j m_j\Pi_{ij}\nabla_iW_{ij},\qquad
 \Pi_{ij}=\begin{cases}
 -\alpha_{\rm AV}c_0\mu_{ij}/\bar\rho_{ij},&\bm v_{ij}\!\cdot\bm r_{ij}<0,\\
 0,&\text{otherwise},
 \end{cases}\label{eq:artificialviscosity}\\
 \mu_{ij}&=\frac{h\,\bm v_{ij}\!\cdot\bm r_{ij}}{|\bm r_{ij}|^2+0.01h^2},\qquad
 \bar\rho_{ij}=\frac{\rho_i+\rho_j}{2},\qquad \bm v_{ij}=\vs_i-\vs_j.
 \nonumber
\end{align}
Here $c_0$ is the numerical reference sound speed; no quadratic artificial-viscosity term is used. Wall-neighbour velocities follow the assigned mechanical slip condition. The value reported as artificial viscosity in the cases is $\alpha_{\rm AV}=0.1$.

Table~\ref{tab:stabilization} gives the damping and artificial-stress settings for the principal simulations. Artificial stress is disabled in the elastic benchmarks and enabled in the sensitive-clay slopes. These choices are kept distinct from the stress-diffusion and pore-pressure regularization settings specified with each case.
\begin{table}[tb]
\centering\small
\caption{Numerical stabilization parameters for the principal simulations. Off indicates disabled artificial stress; the exponent then has no effect. All listed simulations use $\alpha_{\rm AV}=0.1$. The smoothing-length ratio is included because it affects both the damping rate and the artificial-stress interaction.}
\label{tab:stabilization}
\begin{tabular}{llrrrr}
\toprule Case & Stage & $\xi$ & $\epsilon_{\rm AS}$ & $q_{\rm AS}$ & $h/\Delta x$\\\midrule
1D column & Consolidation & 0.02 & Off & --- & 1.8\\
2D strip (50 s) & Loading/drainage & 0.04 & Off & --- & 1.8\\
Cryer sphere & Loading/drainage & 0.02 & Off & --- & 1.8\\
5 m slope & Prestress & 0.02 & 0.2 & 4 & 1.5\\
5 m slope & Failure & $10^{-5}$ & 0.3 & 5 & 2.16\\
8 m slope & Prestress & 0.02 & 0.2 & 4 & 1.5\\
8 m slope & Failure & $10^{-5}$ & 0.2 & 4 & 1.5\\\bottomrule
\end{tabular}
\end{table}

\subsection{Stress diffusion and pore-pressure regularization}\label{sec:regularization}
Two further numerical operations act on the stress and pressure fields. Stress diffusion contributes to the effective-stress rate during force evaluation, whereas Shepard regularization replaces a pressure field at selected completed time steps. Their activation is independent, and neither is identified with physical Darcy drainage.

\paragraph{Stress diffusion.}
The extension retains the stress-diffusion approach developed for geomechanical SPH by \citet{feng2021stress} and used in GeoDualSPHysics \citep{feng2026geodual}. The gravity-corrected soil--soil form selected for the slopes adds the following tensor to the constitutive effective-stress rate:
\begin{align}
 \bm D_{\sigma,i}^{\rm diff}
 &=2\delta_\sigma h c_0\sum_{j\in\mathcal N_i^s}V_j
 \left(\sige_i-\sige_j+\bm H_{ij}\right)
 \frac{\bm r_{ij}\!\cdot\nabla_iW_{ij}}{|\bm r_{ij}|^2+0.01h^2},\label{eq:stressdiffusion}\\
 \bm H_{ij}&=\rho_0 g_z(z_i-z_j)\,
 \mathrm{diag}(K_0,K_0,1),\qquad K_0=1-\sin\phi.\label{eq:stressdiffusionreference}
\end{align}
Here $\delta_\sigma$ is dimensionless, $c_0$ is the numerical reference sound speed, $\rho_0$ is the reference mixture density, and $g_z$ is the signed vertical gravity component. The prefactor follows the implemented normalization $\delta_\sigma(2h)c_0$, using the Wendland support radius $2h$. The sum includes neighbouring deformable-soil particles and excludes rigid-wall and floating-body neighbours in this selected form. It acts on all six stress components; the gravity increment changes only the diagonal ones. The same implementation selector and amplitude are shared with the density-diffusion channel, so these calculations do not isolate stress diffusion from density diffusion.

The gravity increment represents a prescribed reference stress gradient. Because it uses $\rho_0$, it does not by itself establish exact effective-stress equilibrium for an arbitrary saturated hydrostatic state. When enabled, diffusion is assembled at each force evaluation and added before stress integration; it is not a periodic postprocessing filter. Both slope geometries use $\delta_\sigma=0.1$ during prestress and failure. Stress/density diffusion is disabled in the principal 1D, 2D strip, and Cryer benchmarks. Other available diffusion variants are not used for the results reported here.

\paragraph{Pore-pressure Shepard regularization.}
The optional pressure filter operates on the excess field $p_{w,j}^e=\pw_j-p_{w0,j}$ rather than directly averaging total pore pressure. For an eligible soil particle, the replacement is
\begin{equation}
 \pw_i^{\rm new}=p_{w0,i}
 +\frac{\displaystyle\sum_{j\in\mathcal S_i}V_jW_{ij}
 (\pw_j^{\rm old}-p_{w0,j})}
 {\displaystyle\sum_{j\in\mathcal S_i}V_jW_{ij}}.
 \label{eq:poreshepard}
\end{equation}
The support $\mathcal S_i$ includes the particle itself with $W_{ii}=W(0,h)$, valid soil neighbours, and admitted wall neighbours within the kernel support, with $V_j=m_j/\rho_j$ and positive density. Hydraulic mDBC activation rules govern wall admission; for its no-slip/free-slip variants, non-submerged wall particles deactivated by the boundary procedure are omitted. All averages use the pre-filter field and are written to a separate field before replacement. If the denominator is zero, the old pressure is retained. Boundary particles supply reconstructed values but are not themselves updated by the filter.

There is no additional blending coefficient: an enabled filter performs the full normalized replacement in Equation~\eqref{eq:poreshepard}. It requires active hydromechanical coupling, an enabled pressure-regularization option, and a positive integer interval $N_S$. In the predictor--corrector calculations reported here, it is applied after a completed step when the completed-step count is a multiple of $N_S$, excluding predictor-stage neighbour rebuilds. Hydraulic boundary reconstruction is refreshed immediately before and after filtering. Active drained-surface particles retain their prescribed zero gauge pressure, which is re-enforced after replacement; impermeable loaded contacts remain excluded from the drained mask as described in Section~\ref{sec:bc}. The stored reference field $p_{w0}$ is unchanged.

Both slopes use $N_S=40$ in prestress and failure. With $\Delta t=10^{-6}$ s, this is a filtering interval of $4\times10^{-5}$ s. The principal elastic benchmarks do not use this filter. Normalization preserves a constant excess-pressure field over eligible support, but does not impose a discrete global pore-water-mass constraint. The filter can modify pressure gradients and hence subsequent deformation, so its interval requires sensitivity assessment alongside the stress-diffusion strength. Table~\ref{tab:regularization} summarizes the settings.
\begin{table}[tb]
\centering\small
\caption{Stress-diffusion and pore-pressure Shepard settings for the principal simulations. Off denotes a disabled operation. Active stress diffusion uses the gravity-corrected soil--soil form; $N_S$ counts completed time steps.}
\label{tab:regularization}
\begin{tabular}{llrlr}
\toprule Case & Stage & $\delta_\sigma$ & Pressure filter & $N_S$\\\midrule
1D column & Consolidation & Off & Off & ---\\
2D strip (50 s) & Loading/drainage & Off & Off & ---\\
Cryer sphere & Loading/drainage & Off & Off & ---\\
5 m and 8 m slopes & Prestress & 0.1 & On & 40\\
5 m and 8 m slopes & Failure & 0.1 & On & 40\\\bottomrule
\end{tabular}
\end{table}

\subsection{Mechanical and hydraulic boundary treatment}\label{sec:bc}
The boundary representation extends the mDBC infrastructure of GeoDualSPHysics \citep{english2022,feng2026geodual}. Mechanical recovery uses ghost locations inside the material support and local reconstruction of stress and velocity. No-slip and free-slip behaviour can be assigned by boundary marker. For a stationary free-slip wall, normal motion is suppressed while tangential motion is permitted; a fixed no-slip wall suppresses both components.

Hydraulic reconstruction operates on excess pressure separately from effective stress. For the flexible-strip simulations, zero-order interpolation is used and an interior ghost estimate is obtained by volume-weighted interpolation,
\begin{equation}
 p_{w,g}^e=\frac{\sum_j V_jW_{gj}(\pw_j-p_{w0,j})}{\sum_jV_jW_{gj}},
 \qquad \pw_b=p_{w0,b}+p_{w,g}^e .\label{eq:ghost}
\end{equation}
An MLS reconstruction is also available, but the flexible-strip baseline uses the zero-order interpolation in Equation~\eqref{eq:ghost}. The reconstructed boundary pressure supplies the Darcy interaction directly. Separating $p_{w0,b}$ from the excess increment is intended to preserve a hydrostatic reference under gravity rather than interpolate total pressure without accounting for elevation. This is a practical ghost-value closure for an impermeable boundary; exact discrete zero flux and hydrostatic preservation still require verification for each geometry and operator combination.

The physical hydraulic conditions are
\begin{equation}
 \pw=\bar p_w\quad\hbox{on }\Gamma_D,
 \qquad \bm q\!\cdot\bm n=0\quad\hbox{on }\Gamma_N.
\end{equation}
For a gravity-free impermeable wall the second condition is a zero normal pressure gradient. With gravity it is a zero normal gradient of hydraulic head, not generally of pressure alone. Drained free-surface particles are identified by the surface-tracking procedure and assigned the prescribed pressure, zero gauge pressure in the examples here. Load footprints are excluded from this mask where the contact is impermeable. Drainage can be activated after a loading ramp without changing the mechanical support.

\begin{figure}[tb]
\centering\includegraphics[width=\linewidth]{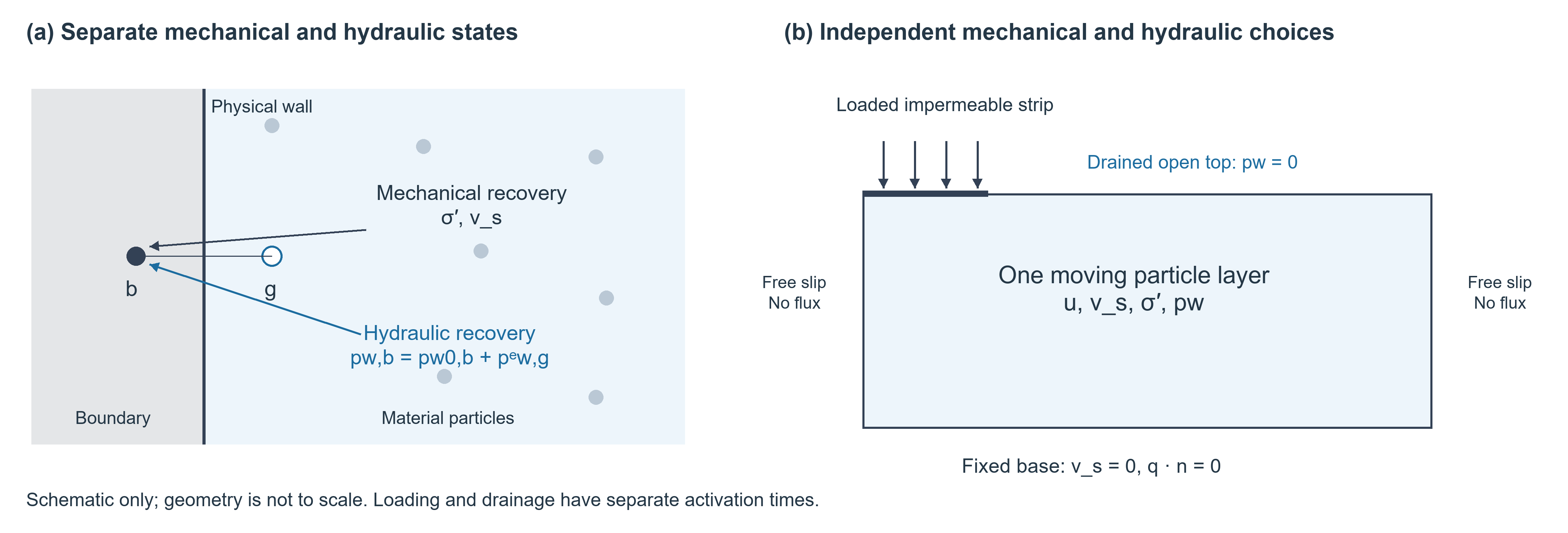}
\caption{Conceptual boundary organization of the hydraulic extension. (a) An interior ghost location $g$ supports separate recovery of mechanical state and excess pressure at boundary particle $b$. (b) Mechanical slip and hydraulic drainage are independent choices; the strip remains impermeable when the neighbouring top surface begins to drain. The drawing is schematic and is not to scale.}
\label{fig:boundary}
\end{figure}

Two traction treatments are distinguished from rigid-wall mDBC. The strip test applies a vertical distributed load over its selected surface footprint. Cryer's sphere uses flexible confinement based on the kernel-truncation approach of \citet{zhao2019}. Its curved surface simultaneously carries confining traction and, after the prescribed activation time, zero pore pressure. Thus Cryer's test assesses the three-dimensional coupled solution and drained curved surface; it is not by itself a test of impermeable rigid-wall mDBC. Figure~\ref{fig:boundary} summarizes the separation of mechanical and hydraulic states.

\begin{table}[tb]
\centering\small
\caption{Mechanical and hydraulic conditions used in the principal tests.}
\label{tab:bc}
\begin{tabular}{>{\raggedright\arraybackslash}p{29mm}>{\raggedright\arraybackslash}p{52mm}>{\raggedright\arraybackslash}p{61mm}}
\toprule Case & Mechanical conditions & Hydraulic conditions\\\midrule
1D column & Laterally periodic column; supported base; top surcharge & Closed during ramp; top drained afterwards; base impermeable\\
2D strip & Free-slip lateral walls; fixed no-slip base; loaded strip & Initially closed; open top drained after 1 s; strip and remaining walls impermeable\\
Cryer sphere & Uniform flexible confining traction & Closed during 0.0025 s ramp; exterior drained afterwards\\
Sensitive-clay slopes & Rigid support; 5 m slope uses a no-slip wall condition & Tracked free surface drained; supported walls use hydraulic ghost reconstruction\\\bottomrule
\end{tabular}
\end{table}

\subsection{Time integration, stabilization, and verification strategy}
Pressure and motion are advanced with the explicit predictor--corrector/symplectic infrastructure. Each force evaluation coordinates surface identification, boundary recovery, mechanical interactions, pressure-rate assembly, and the pressure update. Prescribed drained values are enforced in the relevant stages. Restart states transfer effective stress together with $\pw$ and $p_{w0}$, allowing the failure stage to inherit the prestress solution.

The explicit pressure equation introduces a diffusion-related time scale of order
\begin{equation}
 \Delta t_p\sim C_p\frac{h^2}{c_p},\qquad
 c_p=\frac{K_w k}{n\rho_w g},
\end{equation}
in addition to the mechanical step restrictions. This estimate describes scaling rather than a universal stability bound. The fixed increments actually used are reported below. No time-projection integration or pressure-Poisson solution is claimed for the presented PR results. Artificial stress, artificial viscosity, and damping are defined in Section~\ref{sec:stabilization}; stress diffusion and pressure regularization are defined in Section~\ref{sec:regularization} and specified by case. These numerical treatments are not interpreted as physical pore-fluid dissipation.

The verification strategy progresses from one-dimensional consolidation to two-dimensional strip loading and the three-dimensional Cryer problem, followed by large-deformation slope applications. Quantitative errors are evaluated for the one-dimensional consolidation histories and the four closed-ramp Cryer simulations. Pressure-field snapshots and slope deformation sequences provide complementary qualitative assessments. Differences in time step, mechanical support, and loading history are specified for each comparison.

\begin{table}[tb]
\centering\small
\caption{Material properties and particle spacing for the principal simulations. The short-duration strip-loading variant is specified separately in Section~\ref{sec:strip}. The slope parameters apply to both geometries.}
\label{tab:materials}
\begin{tabular}{lrrrrrr}
\toprule Case&$E$ (MPa)&$\nu$&$n$&$K_w$ (GPa)&$k$ (m/s)&$\Delta x$ (m)\\\midrule
1D column&2&0.30&0.30&0.20&$10^{-4}$--$10^{-3}$&0.01\\
2D full strip&20&0.33&0.40&1.00&$10^{-3}$&0.10\\
Cryer sphere&2&0.10--0.45&0.30&0.20&$10^{-5}$&0.002\\
Sensitive-clay slopes&25&0.30&0.40&0.20&$10^{-8}$&0.10\\\bottomrule
\end{tabular}
\end{table}

\section{Numerical tests}\label{sec:tests}
\subsection{One-dimensional consolidation}
\begin{figure}[tb]
\centering\includegraphics[width=0.76\linewidth]{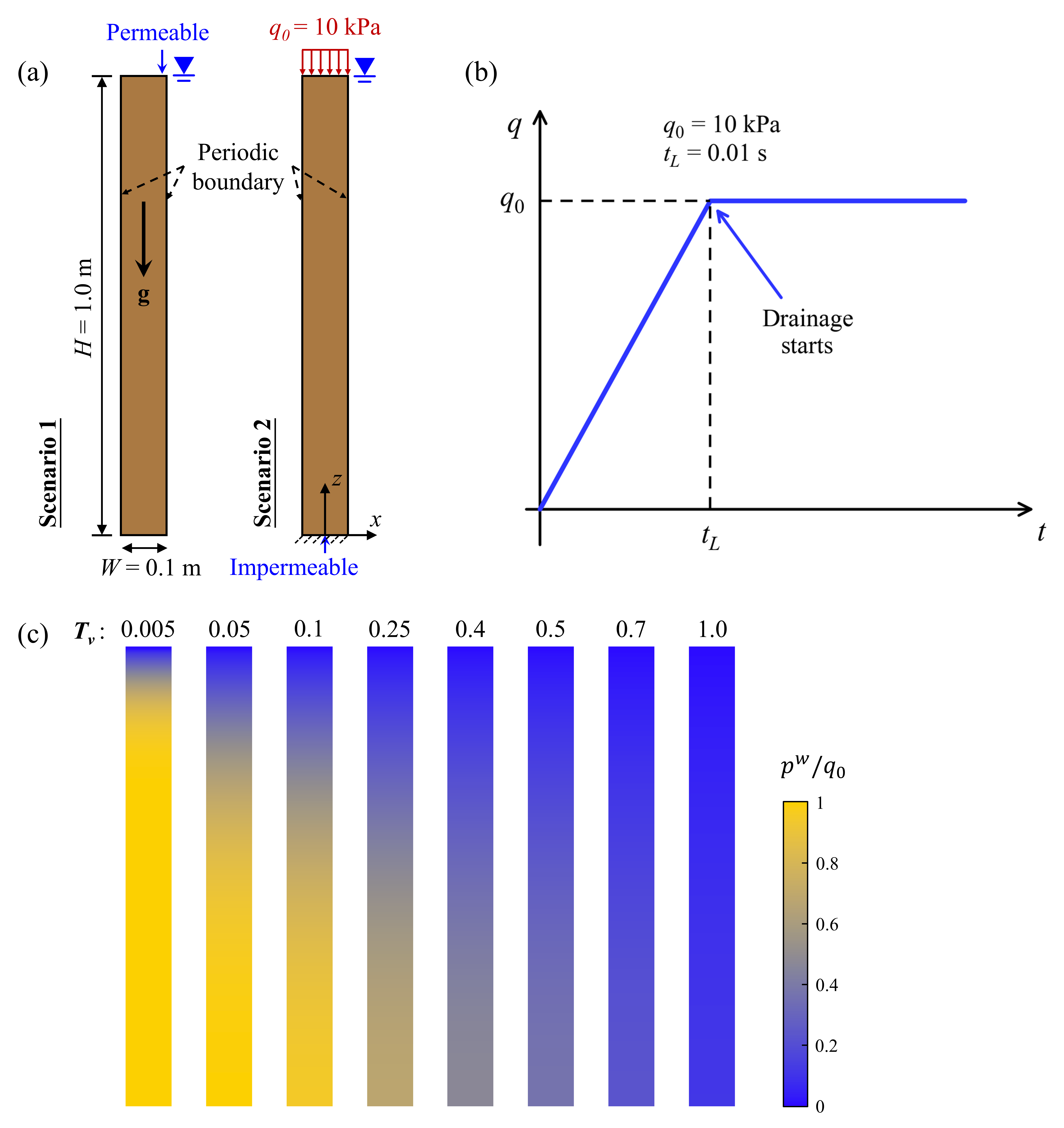}
\caption{Schematic of the 1D column, surcharge ramp, drainage activation, and illustrative pressure dissipation. The quantitative results reported in the text concern the external-loading scenario. The self-weight scenario is shown for context; the quantitative assessment concerns the external-loading case.}
\label{fig:columnsetup}
\end{figure}

The external-loading column follows the benchmark discussed by \citet{castillo2026}, with the classical consolidation solution of \citet{terzaghi1943} as reference. The column is $H=1$ m high and 0.1 m wide, with initial spacing $\Delta x=0.01$ m, lateral periodicity, zero gravity, and a surcharge $q_0=10$ kPa. The material parameters are $E=2$ MPa, $\nu=0.3$, $n=0.3$, $K_w=0.2$ GPa, $\rho=2100$ kg/m$^3$, and $\rho_w=1000$ kg/m$^3$. A 0.01 s ramp precedes top drainage for the three time-history comparisons. Figure~\ref{fig:columnsetup} shows the loading and drainage sequence.

For the classical one-way-drained reference, define the constrained modulus $M_c=E(1-\nu)/[(1+\nu)(1-2\nu)]$, $c_v=kM_c/(\rho_w g)$, and $T_v=c_v(t-t_L)/H^2$ after drainage starts. For depth $y$ measured downwards from the drained top,
\begin{align}
 \frac{p_w^e(y,T_v)}{q_0}&=\frac{4}{\pi}\sum_{m=0}^{\infty}
 \frac{\sin[(2m+1)\pi y/(2H)]}{2m+1}
 \exp\!\left[-\frac{(2m+1)^2\pi^2T_v}{4}\right],\\
 U(T_v)&=1-\frac{8}{\pi^2}\sum_{m=0}^{\infty}
 \frac{\exp[-(2m+1)^2\pi^2T_v/4]}{(2m+1)^2},\qquad
 S(T_v)=\frac{q_0H}{M_c}U(T_v).
\end{align}
The classical $c_v$ is used to normalize the consolidation histories and reference solutions. The finite-fluid-compressibility equation instead gives a quasi-static storage correction $c_v^*=c_v/[1+nM_c/K_w]$. For this column, $nM_c/K_w\simeq0.0040$; it is small but should be distinguished from discretization error. Finite ramp duration, dynamic settling, and particle-based pressure averaging also affect early-time comparisons \citep{yao2025}.

\begin{figure}[tb]
\centering\includegraphics[width=0.83\linewidth]{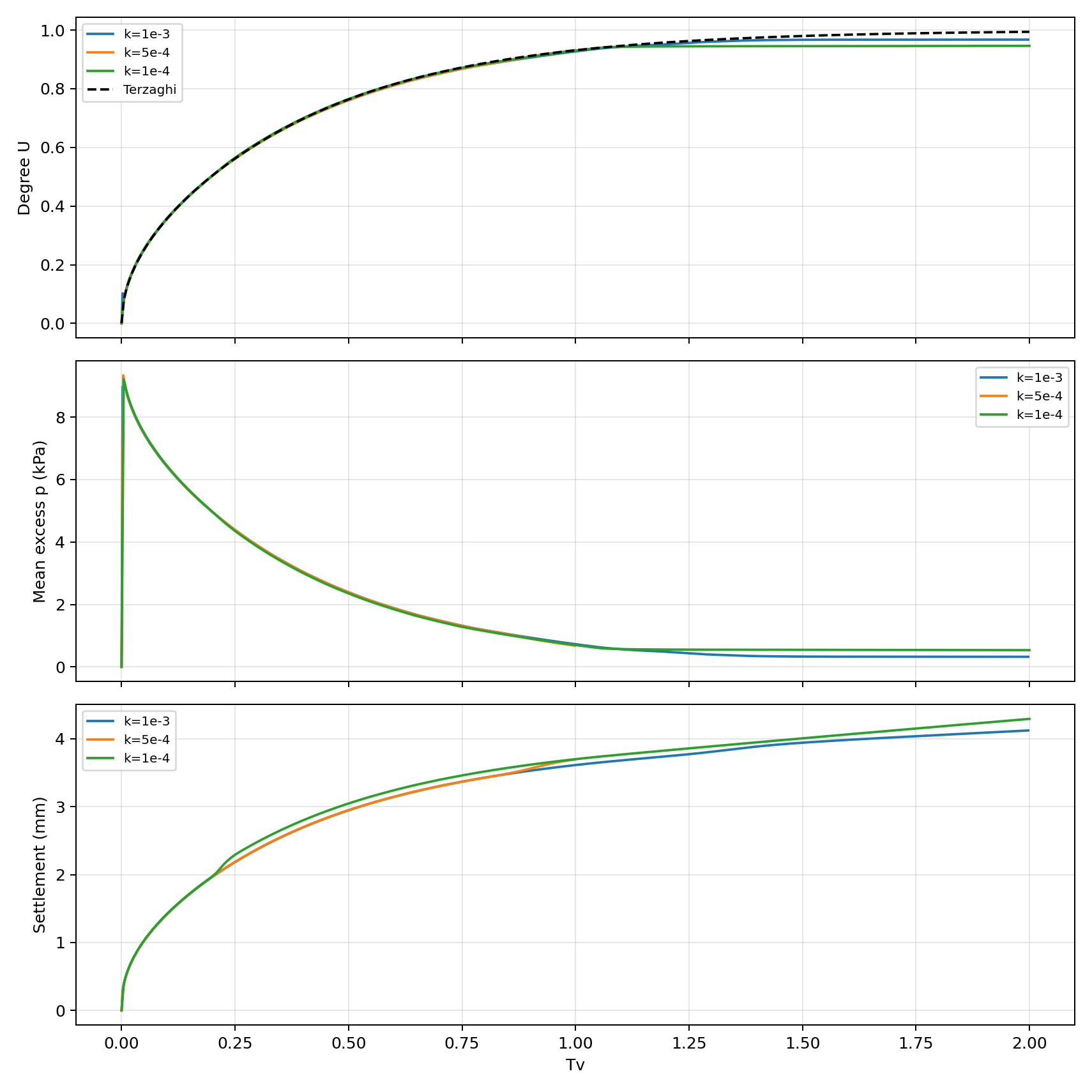}
\caption{One-dimensional external-load consolidation histories: degree of consolidation (top), mean excess pore pressure (middle), and settlement (bottom), with their reference curves. Hydraulic conductivities are $10^{-3}$, $5\times10^{-4}$, and $10^{-4}$ m/s. The intermediate-conductivity series stops near $T_v=1$.}
\label{fig:columnhistory}
\end{figure}

\begin{figure}[!htbp]
\centering\includegraphics[width=0.9\linewidth]{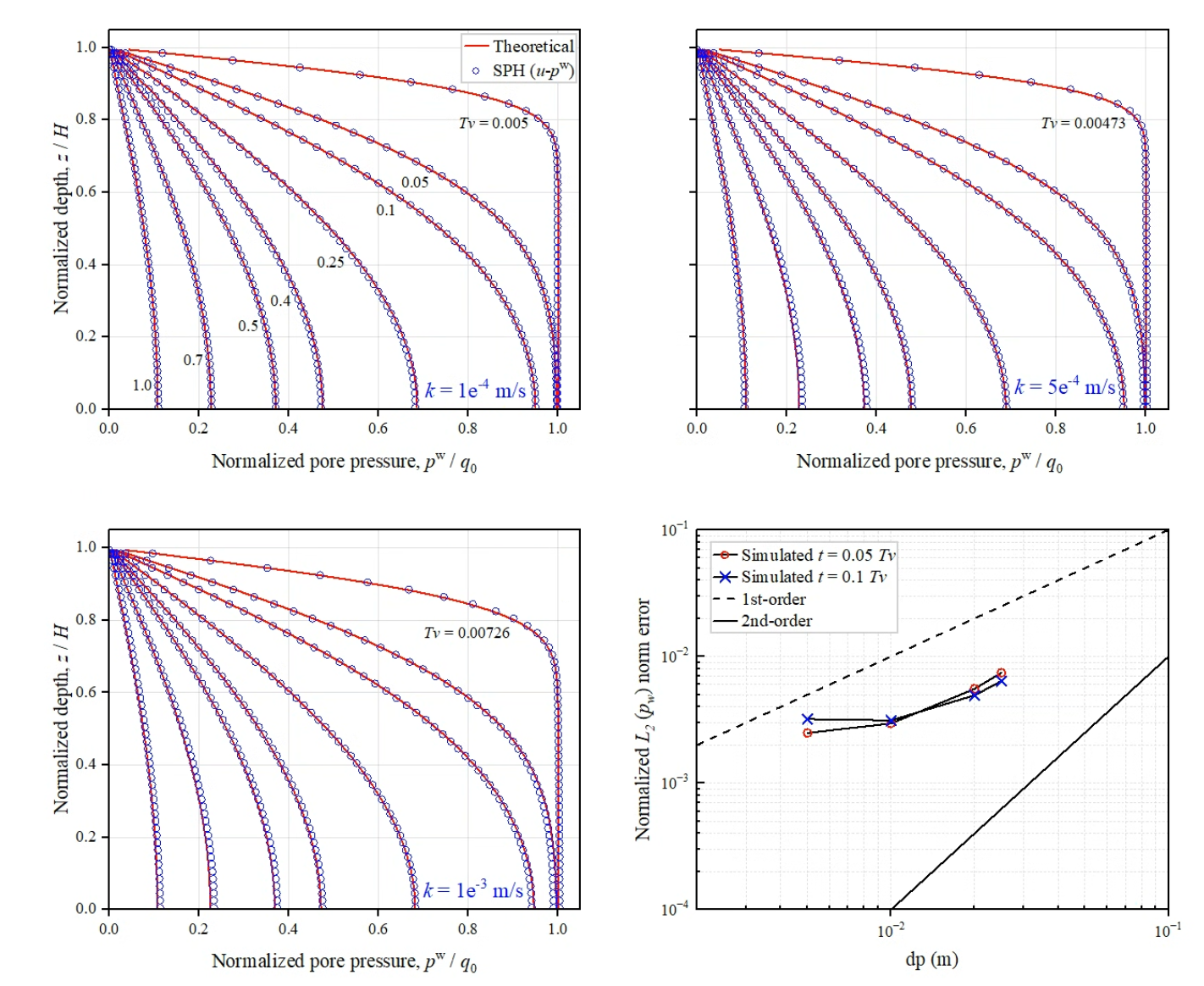}
\caption{One-dimensional consolidation pressure profiles and particle-spacing comparison. The panels compare theoretical and SPH pressure distributions. The spacing comparison is qualitative and is not used to estimate a convergence order.}
\label{fig:profiles}
\end{figure}

Figure~\ref{fig:columnhistory} shows the degree of consolidation, mean excess pressure, and settlement. All series show the expected pressure dissipation and increasing settlement. The intermediate-conductivity series ends near $T_v=1$, whereas the other two extend near $T_v=2$. Table~\ref{tab:column} summarizes the results at the final simulated time of each series. Pressure-based consolidation is closer to the reference than settlement: the final settlement exceeds the reference by approximately 6.8--16.3\%. This difference prevents a claim of uniformly accurate consolidation even where the pressure curves appear close. Figure~\ref{fig:profiles} complements the time histories with depthwise pressure profiles. The family of profiles illustrates the progression of dissipation from the drained surface into the column, with pressure retained for longer near the impermeable base. The particle-spacing comparison is qualitative; a systematic refinement study with matched parameters is required to determine the convergence order.

\begin{table}[tb]
\centering\small
\caption{Summary of the one-dimensional consolidation results. $E_U$ is the root-mean-square error of the pressure-based degree of consolidation. Settlement is evaluated at the last stored time, not at a common $T_v$.}
\label{tab:column}
\begin{tabular}{rrrrrrr}
\toprule $k$ (m/s)&$\Delta t$ (s)&Last $T_v$&$U_{\rm SPH}$&$E_U$&$S_{\rm SPH}$ (mm)&$S_{\rm ref}$ (mm)\\\midrule
$10^{-3}$&$10^{-6}$&1.9973&0.9676&0.01182&4.123&3.692\\
$5\times10^{-4}$&$2\times10^{-6}$&0.9986&0.9319&0.00271&3.694&3.458\\
$10^{-4}$&$10^{-5}$&1.9997&0.9463&0.02383&4.294&3.693\\\bottomrule
\end{tabular}
\end{table}
Artificial viscosity uses $\alpha_{\rm AV}=0.1$ and damping uses $\xi=0.02$ as defined in Section~\ref{sec:stabilization}. Artificial stress, pressure regularization, and stress/density diffusion are disabled in this comparison. Different increments are used for the three conductivities; these calculations do not constitute a systematic time-step convergence study.

\FloatBarrier

\subsection{Two-dimensional consolidation under a flexible strip}\label{sec:strip}
The two-dimensional benchmark follows the flexible-strip test of \citet{lian2023}, Section 4.3 and Figures 9--11. The computational layer is 20 m wide and 10 m deep, with a 1.25 m loaded strip on the left part of the top surface and $\Delta x=0.1$ m. The load reaches 10 kPa at $t_L=1$ s and remains constant. All boundaries are impermeable during the ramp. Afterwards, the open top drains while the strip contact remains impermeable. The side walls are free-slip and the base is fixed in both directions through separate mDBC marker assignments.

The parameters are $E=20$ MPa, $\nu=0.33$, $n=0.4$, $K_w=1$ GPa, $k=10^{-3}$ m/s, and zero gravity. The mixture density is 1990 kg/m$^3$, consistent with $\rho_s=2650$ kg/m$^3$ and $\rho_w=1000$ kg/m$^3$ at this porosity. The full run uses $\Delta t=10^{-5}$ s, damping coefficient $\xi=0.04$, artificial viscosity $\alpha_{\rm AV}=0.1$, no artificial stress, no pore-pressure regularization, and no stress/density diffusion. The initial particle count is 21,945 including 20,301 material particles.

\begin{figure}[tb]
\centering\includegraphics[width=0.88\linewidth]{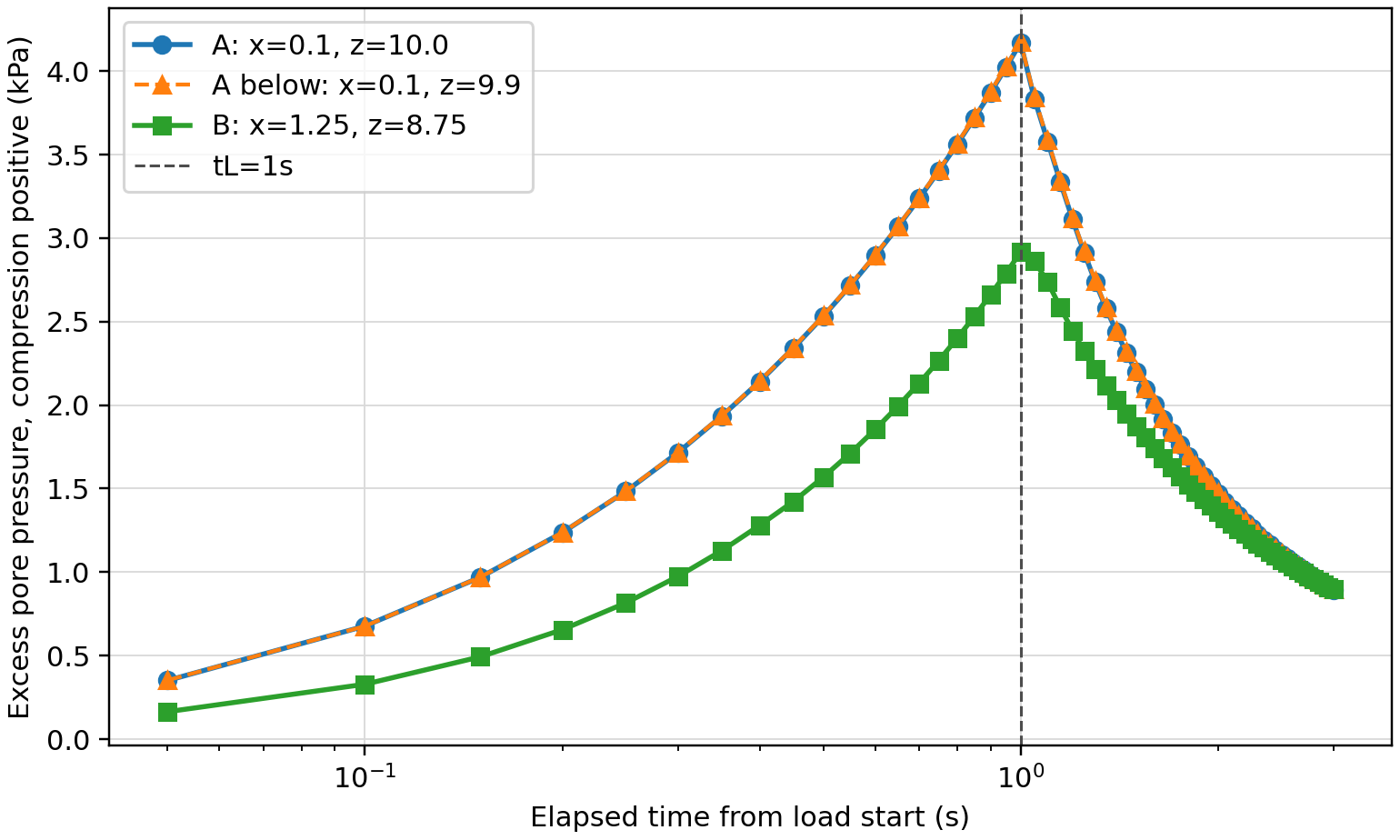}
\caption{Short-duration strip-loading pressure histories. A is at $(x,z)=(0.1,10)$ m, the adjacent subsurface point at $(0.1,9.9)$ m, and B at $(1.25,8.75)$ m. Time is measured from load onset and the vertical marker denotes the 1 s ramp completion. This 3 s calculation uses a different time step and wall configuration from the 50 s pressure-field calculation.}
\label{fig:stripprobes}
\end{figure}

Figure~\ref{fig:stripprobes} presents a separate short-duration calculation extending to 3 s, with $\Delta t=5\times10^{-5}$ s and free-slip conditions on all walls. The 50 s calculation instead uses free-slip lateral walls and a fixed base. The two calculations are considered separately because their time steps and mechanical supports differ. Pressure grows during loading and reaches approximately 4.17 kPa at A and 2.92 kPa at B near $t=1$ s, followed by dissipation. A and the point immediately below it have similar histories, while the deeper point B has a lower loading peak. These curves illustrate compression--pressure feedback and the drainage switch. No quantitative error against the reference time histories is assigned to these curves, and they are not combined with the 50 s calculation into a single time series.

\begin{figure}[!htbp]
\centering\includegraphics[width=\linewidth]{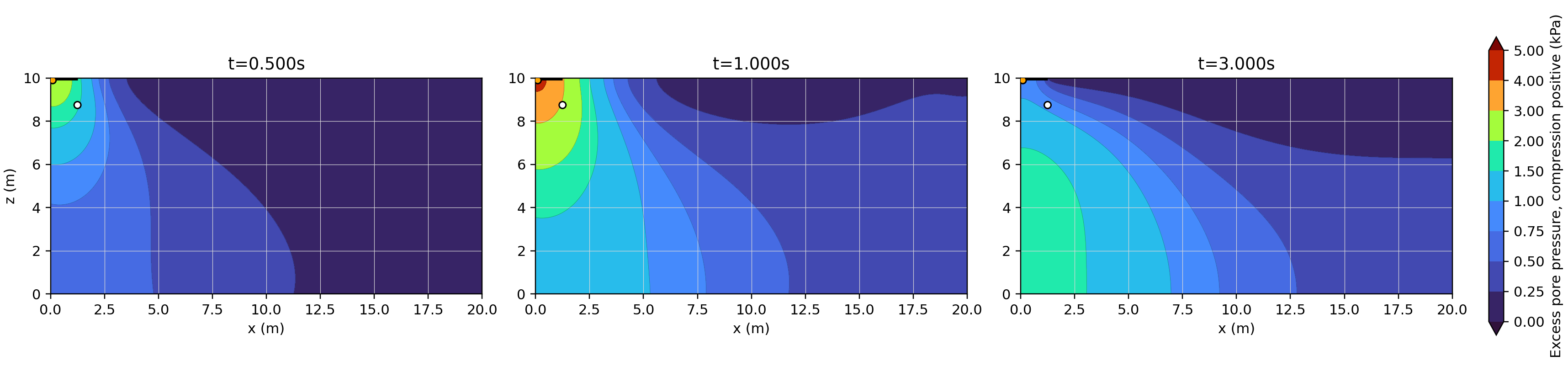}
\caption{Short-duration strip-loading pressure contours at 0.5, 1, and 3 s, with a common colour scale. These fields correspond to the calculation in Figure~\ref{fig:stripprobes}; the 50 s calculation in Figure~\ref{fig:stripfull} uses different mechanical support conditions.}
\label{fig:stripearly}
\end{figure}

Figure~\ref{fig:stripearly} provides the corresponding short-duration pressure fields at 0.5, 1, and 3 s on a common colour scale. Pressure concentrates beneath the loaded strip during the ramp, then decreases near the newly drained open surface while spreading into the layer. The nonzero pressure retained beneath the impermeable contact distinguishes it from the adjacent draining surface. These fields correspond to the short-duration point histories and are assessed separately from the 50 s calculation with mixed mechanical supports.

\begin{figure}[p]
\centering\includegraphics[height=0.74\textheight,trim=330bp 0 0 0,clip]{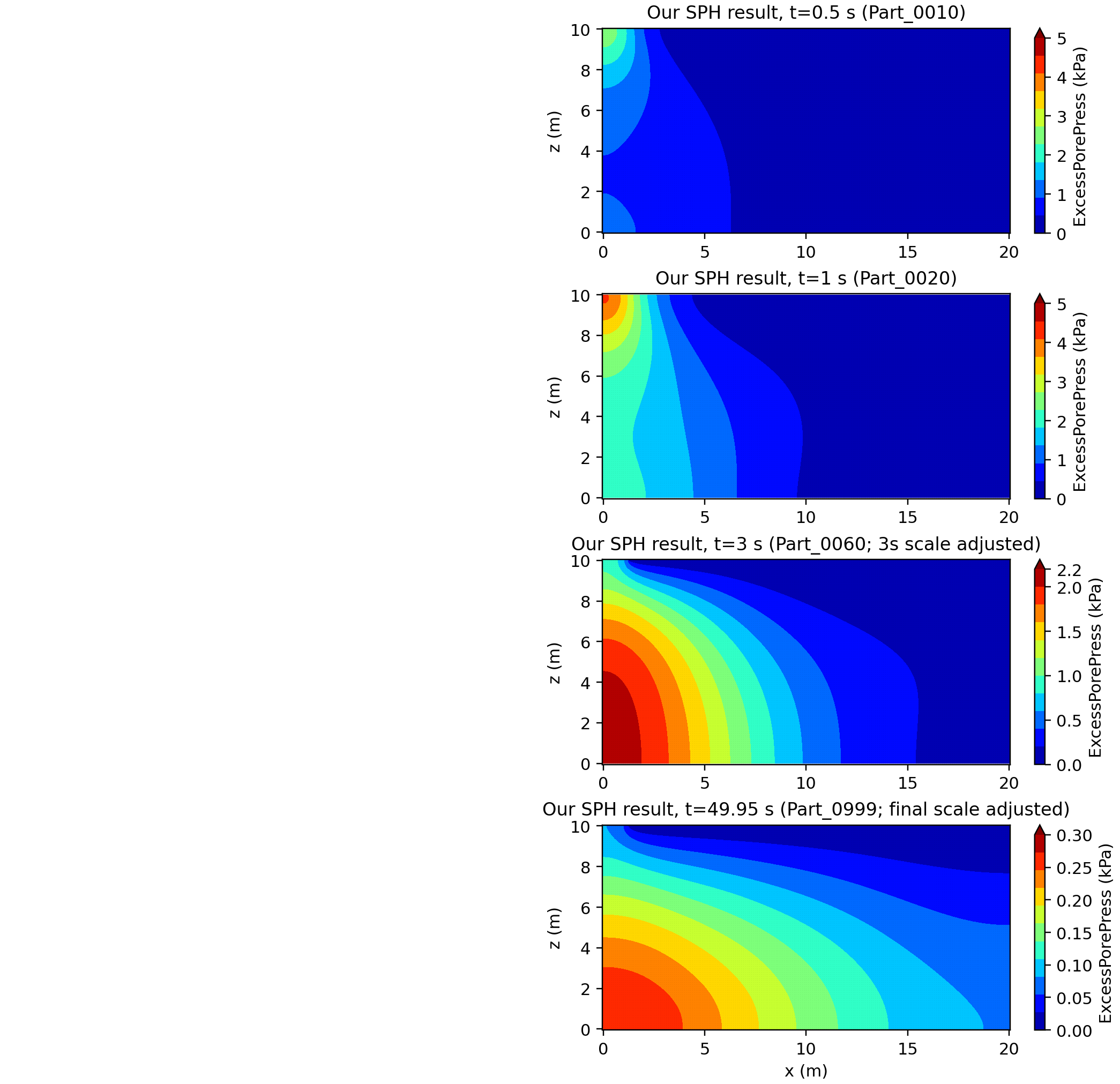}
\caption{Two-dimensional flexible-strip pressure fields from the 50 s simulation, corresponding to the setup of Lian et al.~\citep{lian2023}, Figure 10. Times are approximately 0.5, 1, 3, and 49.95 s from load onset. Independent colour limits are retained: the 3 s panel uses an upper limit of 2.2 kPa and the final panel 0.3 kPa. Equal colours across panels therefore do not represent equal pressures. All panels show the present GeoDualSPHysics simulations.}
\label{fig:stripfull}
\end{figure}

Figure~\ref{fig:stripfull} presents the 50 s calculation at 0.5, 1, 3, and 49.95 s. The footprint remains pressurized after drainage starts outside the loaded contact. The field subsequently spreads into the layer while pressure decreases. The maximum pressures are 2.510, 4.139, 2.118, and 0.269 kPa; the corresponding means are 0.335, 0.695, 0.626, and 0.108 kPa. These values are more informative for comparing amplitudes than the changing colours because each panel has its own scale.

The late field is not fully dissipated at the final displayed time. Late-time pressure dissipation is slower than in the reference results of \citet{lian2023}. A separate 50 s calculation with plastic correction disabled gives the same particle solution as the baseline, indicating that plastic yielding does not explain the residual in this comparison. Short-duration sensitivity calculations show little change from halving the increment from $10^{-5}$ to $5\times10^{-6}$ s, but a stronger sensitivity to damping. These observations locate an unresolved accuracy issue in the elastic hydromechanical calculation; they do not establish a unique cause. The test demonstrates coordinated loading, drainage activation, and mixed mechanical support, while the late-time discrepancy requires further investigation.

\FloatBarrier

\subsection{Three-dimensional Cryer's problem}\label{sec:cryer}
Cryer's sphere provides a more discriminating coupling test than monotonic 1D diffusion. Under uniform surface compression and drained exterior conditions, pressure at the centre can first rise above its initial level before dissipating, owing to the interaction between drainage and skeleton deformation \citep{cryer1963,castillo2026}. This Mandel--Cryer effect cannot be demonstrated by matching a pure diffusion equation alone.

The four closed-ramp simulations use radius $R=0.05$ m, $\Delta x=0.002$ m, $E=2$ MPa, $n=0.3$, $K_w=0.2$ GPa, $k=10^{-5}$ m/s, and $p_0=10$ kPa. Poisson ratios are 0.1, 0.2, 0.3, and 0.45. Flexible confinement is ramped over 0.0025 s while the exterior is closed; exterior drainage begins at ramp completion. The fixed step is $5\times10^{-6}$ s, damping uses $\xi=0.02$, artificial viscosity uses $\alpha_{\rm AV}=0.1$, and artificial stress, pressure regularization, and stress/density diffusion are disabled. Unlike the instantaneous-loading idealization, this numerical loading has a finite initial ramp.

Postprocessing uses $T_v=c_v(t-0.0025)/R^2$ with the Poisson-dependent constrained-modulus coefficient defined above and central sampling within $r\leq\Delta x$. This normalization uses the classical constrained-modulus coefficient. The finite-storage correction is small for the lower Poisson ratios but reaches approximately 1.14\% at $\nu=0.45$; it should be included in a refined error attribution. The analytical reference is the Cryer series used in the benchmark presentation of \citet{castillo2026}.

\begin{figure}[tb]
\centering\includegraphics[width=0.86\linewidth]{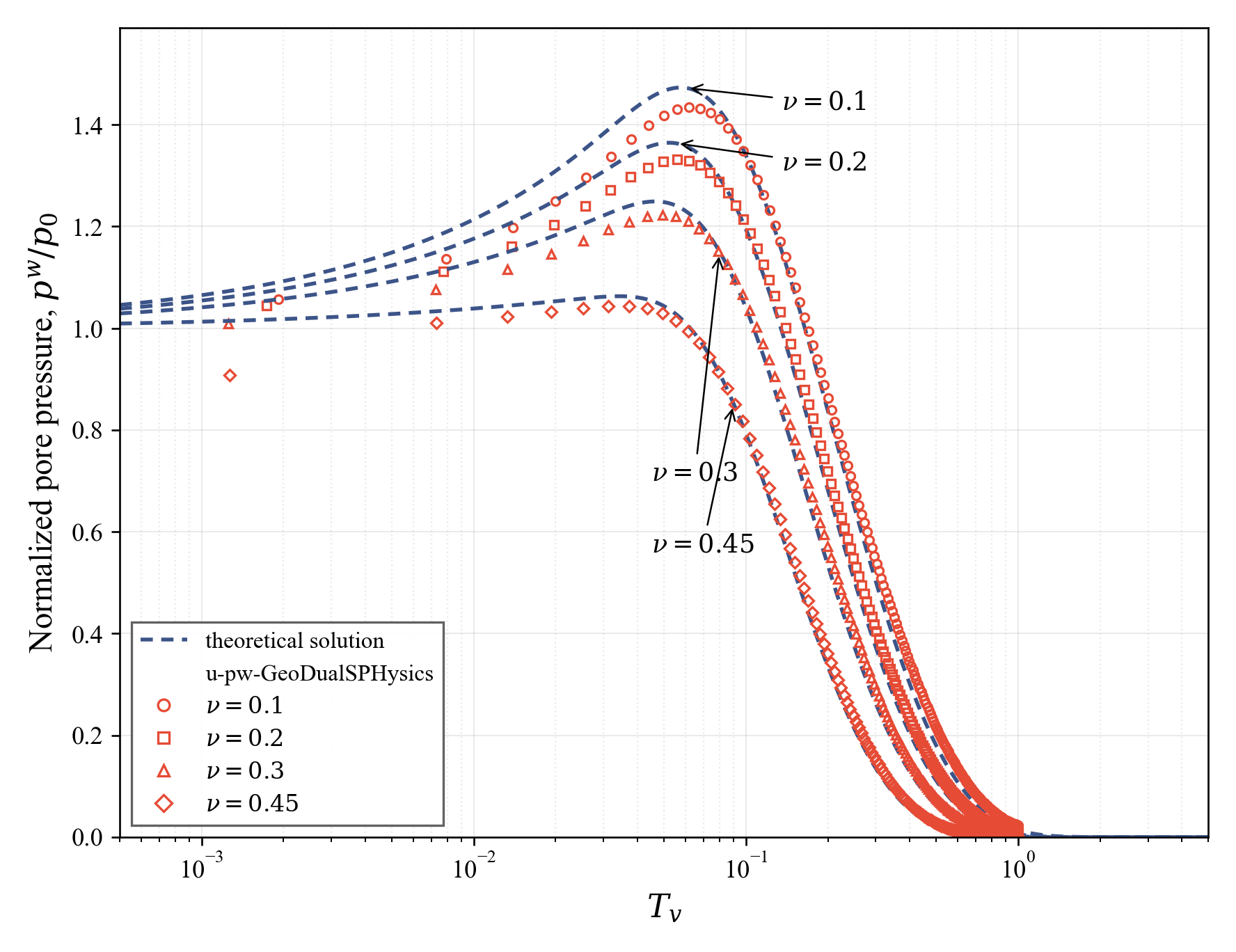}
\caption{Normalized central pore pressure in the closed-ramp Cryer sweep for four Poisson ratios. Symbols show GeoDualSPHysics values sampled within $r\leq0.002$ m; lines show the analytical reference. $T_v$ is measured from drainage activation. The pressure overshoot decreases as the skeleton approaches incompressibility.}
\label{fig:cryer}
\end{figure}

\begin{figure}[!htbp]
\centering\includegraphics[width=\linewidth]{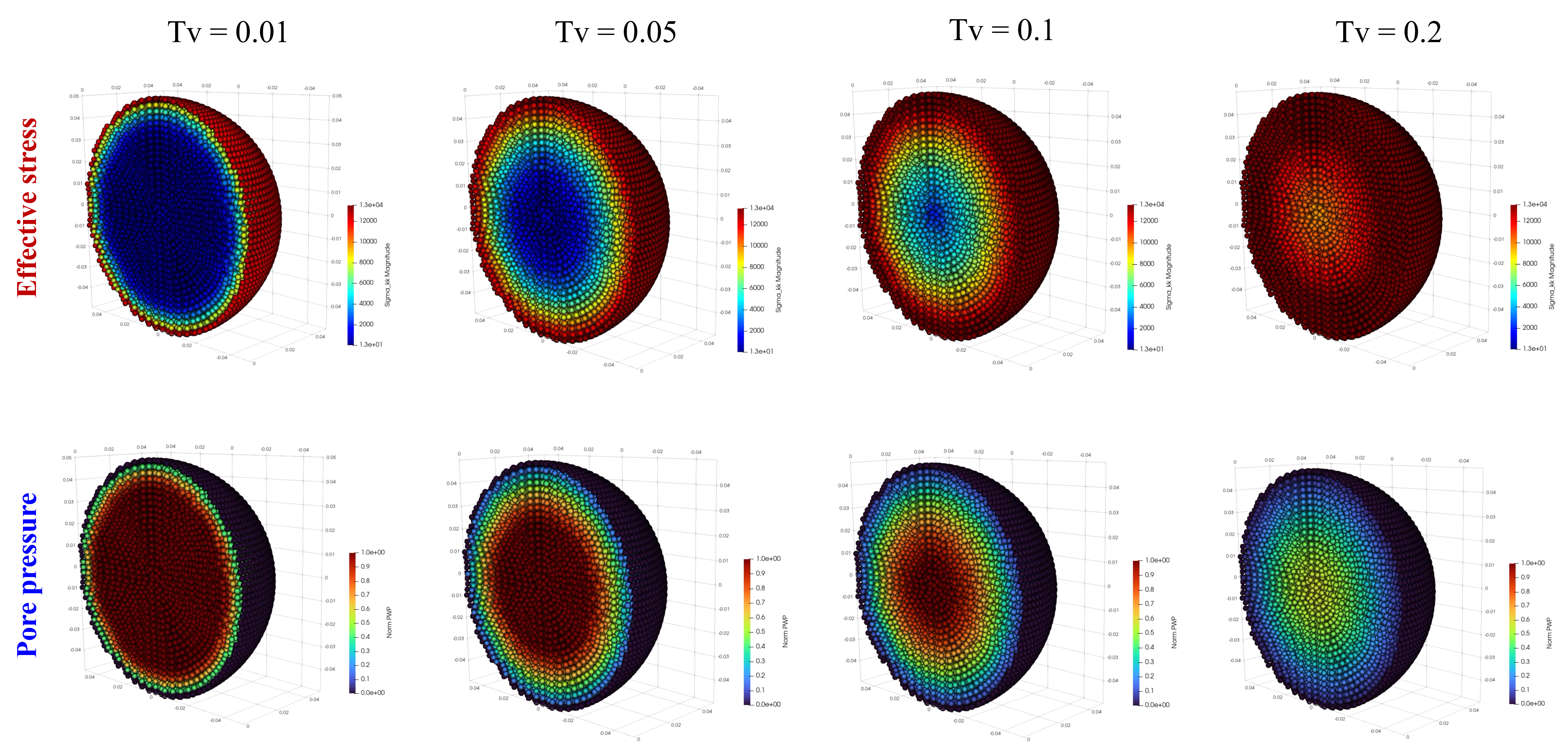}
\caption{Spatial evolution in Cryer's problem at nominal times $T_v=0.01$, 0.05, 0.1, and 0.2, from left to right. The upper row shows effective-stress magnitude and the lower row normalized pore pressure. These qualitative snapshots are assessed separately from the closed-ramp pressure histories and are not included in their error statistics.}
\label{fig:cryerfields}
\end{figure}

Figure~\ref{fig:cryer} reproduces the central-pressure comparison and Table~\ref{tab:cryer} summarizes the corresponding error metrics. The simulations capture the increase and subsequent decay for all four Poisson ratios. Smaller $\nu$ gives a larger overshoot, consistent with poroelastic coupling. Numerical peak pressures are approximately 1.8--2.6\% below the theoretical maxima, and peak occurrence is slightly delayed for the first three groups. The normalized full-series RMSE ranges from 0.0120 to 0.0295. The finite loading ramp, centre sampling, and residual discretization errors all contribute to the comparison; the results should not be described as an exact reproduction of the instantaneous solution. The spatial redistribution of stress and pressure is examined in Figure~\ref{fig:cryerfields} alongside the central-pressure comparison.

\begin{table}[!hb]
\centering\small
\caption{Cryer pressure metrics for the four closed-ramp simulations. Peaks are normalized by $p_0$ and RMSE is computed on normalized pressure. $T_{v,p}$ is the time of the respective maximum.}
\label{tab:cryer}
\begin{tabular}{rrrrrr}
\toprule $\nu$&Peak SPH&Peak reference&$T_{v,p}^{\rm SPH}$&$T_{v,p}^{\rm ref}$&RMSE\\\midrule
0.10&1.4344&1.4728&0.06291&0.05753&0.02953\\
0.20&1.3315&1.3643&0.05673&0.05219&0.02409\\
0.30&1.2226&1.2490&0.04926&0.04636&0.01904\\
0.45&1.0434&1.0628&0.03427&0.03420&0.01201\\\bottomrule
\end{tabular}
\end{table}
Figure~\ref{fig:cryerfields} illustrates the transfer of load towards the skeleton as drainage proceeds. The effective-stress magnitude is initially larger near the exterior and becomes more widely mobilized, while the normalized pore-pressure field decreases as the drained region extends inwards. The snapshots represent nominal times $T_v=0.01$, 0.05, 0.1, and 0.2. They are assessed qualitatively and separately from the closed-ramp central-pressure histories; the error metrics in Table~\ref{tab:cryer} are calculated from those histories alone.

\FloatBarrier

\section{Validation and application: retrogressive landslides}\label{sec:slopes}
\subsection{Slope geometries, initialization, and strength reduction}
The large-deformation assessment adopts the sensitive-clay slope geometries examined by \citet{castillo2026}. The first slope is 5 m high, with a 20 m crest length and 25 m base length, giving a $45^\circ$ face. The second is 8 m high, with 16 m crest and 17 m base lengths, giving a steeper face. Both geometries use a 50 m computational length and $\Delta x=0.1$ m. The two slopes are simulated as separate applications.

Material settings are $\rho=2150$ kg/m$^3$, $E=25$ MPa, $\nu=0.3$, $n=0.4$, $K_w=0.2$ GPa, $\rho_w=1000$ kg/m$^3$, and $k=10^{-8}$ m/s. The low conductivity produces a quasi-undrained interior response over the failure time scale, while the tracked exterior can still carry a drained condition. Friction and dilatancy angles are zero. This idealized pressure-independent softening response captures sensitive-clay strength loss but is not a critical-state or fabric-evolution model.

The calculation separates gravitational prestress from failure. The prestress stage uses peak cohesion $c_{p,0}=25$ kPa, hydrostatic initialization based on the tracked surface, and damping coefficient $\xi=0.02$. The subsequent restart inherits stress and pressure fields and activates softening. At the start of the failure stage, the peak cohesion is reduced by $F_s=1.65$, so the active initial failure-stage cohesion is
\begin{equation}
 c_{p,f}=\frac{25}{1.65}=15.15\ \mathrm{kPa},\qquad
 c(\kappa)=c_r+(c_{p,f}-c_r)\exp(-\eta\kappa),
 \label{eq:softening}
\end{equation}
with residual cohesion $c_r=1.5$ kPa, $\eta=5$, and accumulated plastic measure $\kappa$. Thus 25 kPa is the initialization strength, not the peak used throughout failure. The reduced value is close to the 15.1 kPa reference value, although the stress initialization and boundary implementation remain different from the reference calculation.

The fixed increment is $10^{-6}$ s. Failure-stage damping uses $\xi=10^{-5}$ and artificial viscosity uses $\alpha_{\rm AV}=0.1$. Both slopes use artificial-stress stabilization, gravity-corrected soil--soil stress diffusion with $\delta_\sigma=0.1$, and excess-pressure Shepard regularization with $N_S=40$ during both prestress and failure, as defined in Section~\ref{sec:regularization}. As summarized in Table~\ref{tab:stabilization}, the 5 m slope changes from $(\epsilon_{\rm AS},q_{\rm AS})=(0.2,4)$ during prestress to $(0.3,5)$ during failure, while $h/\Delta x$ changes from 1.5 to 2.16. The 8 m slope uses $(0.2,4)$ and $h/\Delta x=1.5$ in both stages. These stage-dependent numerical choices must be included when interpreting localization and quantitative differences between the geometries.

\subsection{Failure sequence and large deformation}
\begin{figure}[tb]
\centering\includegraphics[width=\linewidth]{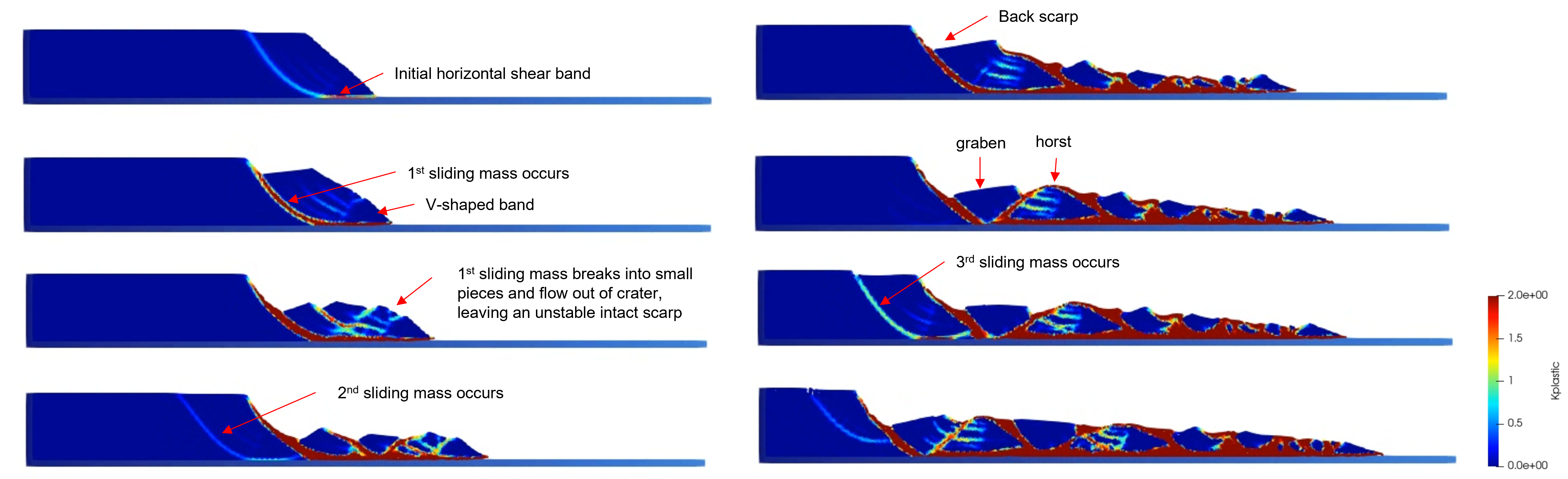}
\caption{Five-metre retrogressive-slope sequence. Colours show the accumulated plastic measure $\kappa$. Read down the left column and then down the right column. The sequence illustrates initial localization, the first sliding mass, subsequent scarp retreat, and further sliding. The panels depict successive deformation stages; quantitative failure times are not evaluated.}
\label{fig:slope5}
\end{figure}

Figure~\ref{fig:slope5} shows the 5 m failure sequence. Localization develops near the toe and into the initial sliding surface. The first moving mass loses support, fragments, and travels downslope, leaving an exposed scarp. Further deformation behind that scarp produces subsequent sliding masses and progressive upslope retreat. The annotated sequence identifies a V-shaped band, successive blocks, and horst--graben features. The SPH material description follows these transitions through separation and redistribution without a fixed mesh connectivity.

The colours represent the accumulated plastic measure $\kappa$, which identifies deformation localization rather than pore-water pressure. The sequence therefore supports the interpretation of localization and retrogression but cannot, by itself, establish where positive or negative excess pressure developed. The inherited pressure fields and explicit pressure evolution provide the hydromechanical calculation, while a direct pressure-versus-time comparison is still required to quantify its influence on the failure sequence.

\begin{figure}[tb]
\centering\includegraphics[width=\linewidth]{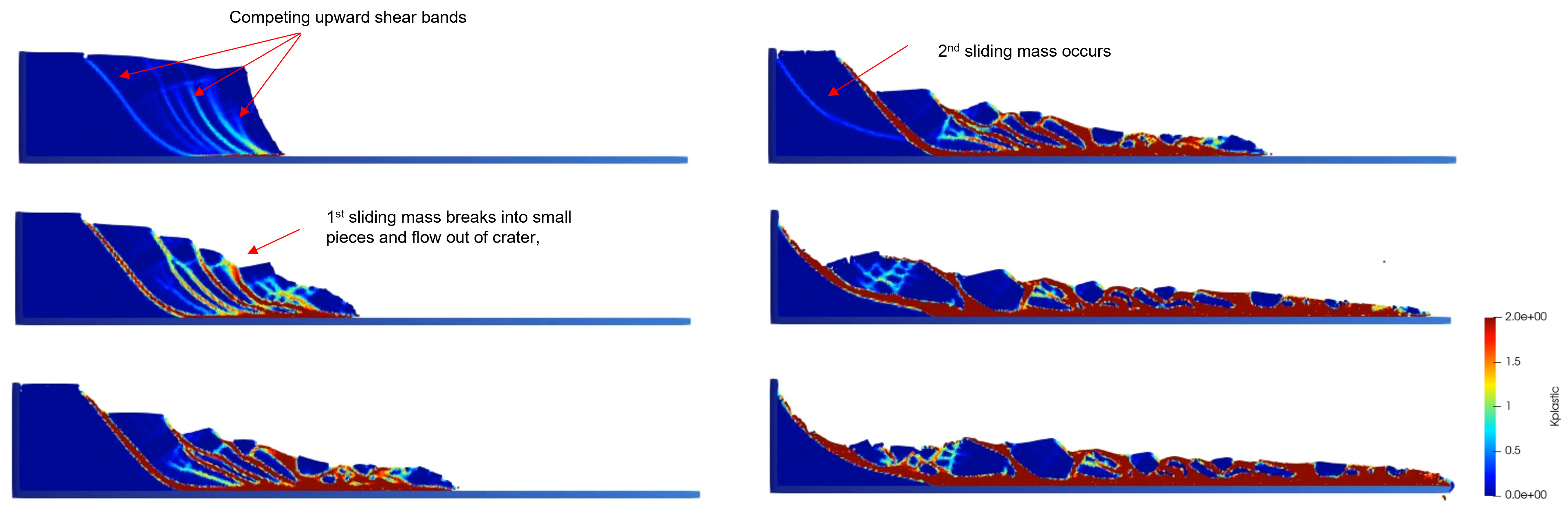}
\caption{Eight-metre retrogressive-slope sequence, coloured by the accumulated plastic measure $\kappa$. Read down the left column and then down the right column. The panels illustrate failure morphology; quantitative initiation times and runout distances are not evaluated.}
\label{fig:slope8}
\end{figure}

The 8 m sequence in Figure~\ref{fig:slope8} shows a steeper initial configuration followed by block release, fragmentation, and repeated retrogression. It extends the qualitative assessment to a second geometry. The assessment of the 8 m case is limited to the deformation patterns shown. Initiation times, runout distances, and quantitative error measures are not evaluated. The sequence therefore illustrates the failure mechanism for this geometry without establishing quantitative predictive accuracy.

These examples establish that the extension can be exercised in a large-deformation workflow with evolving surfaces and repeated failure. They do not yet isolate the effect of hydraulic feedback relative to a penalty-based undrained calculation. A matched comparison would require the same initial stress field, reduced strength, smoothing length, and stabilization in both formulations, as well as pressure histories and geometrical measures. The reference study provides a basis for that comparison \citep{castillo2026}; the present images support a more limited qualitative assessment.
\FloatBarrier

\section{Discussion}
The results identify two complementary roles for the hydraulic extension. First, the pressure-rate equation provides the physical feedback needed for compression-induced pressure generation and drainage. The Cryer overshoot is the clearest evidence that the numerical response includes this interaction beyond a prescribed pressure field or uncoupled diffusion. Second, the mDBC extension places hydraulic states alongside the existing mechanical recovery. The flexible-strip test exercises free-slip sides, a fixed base, an impermeable loaded contact, and an adjacent surface whose drainage changes in time. These boundary distinctions are central to using the solver beyond a single idealized consolidation test.

The evidence also sets limits on accuracy claims. The 1D histories display the expected trend but retain settlement excess and late pressure residuals. The 2D full run preserves the intended drainage mask and load footprint, yet its remaining pressure at long time prevents a claim of complete agreement with the reference solution. Damping, residual compression sources, and hydraulic ghost reconstruction should be examined together because pressure generation scales with $K_w/n$. The existence of an MLS option or a smaller-step diagnostic does not establish that the boundary treatment is fully converged.

The slopes add a different limitation: qualitative agreement of failure morphology is not independent physical validation. Softening localization can depend on spacing, support radius, and artificial-stress parameters. The changes in smoothing length and artificial-stress amplitude and exponent between prestress and failure in the 5 m run should be removed or assessed before attributing differences in retrogression solely to pressure feedback. Damping also influences transient motion and the compression source of pore pressure, even though it does not directly enter the Darcy operator. Quantitative assessment of the 8 m case also requires time-resolved pressure and displacement histories, together with measures of initiation, retrogression, and runout. None of these issues negates the observed deformation sequence, but each constrains the strength of the conclusion drawn from it.

Finally, the extension retains the CPU/GPU architecture of GeoDualSPHysics, but inherited parallel execution does not itself demonstrate a speedup for the coupled method. A performance comparison would need matched hardware, particle count, accuracy, physical duration, and pressure formulation. No such speedup is claimed here. The explicit diffusion restriction remains a practical issue for high conductivity and large fluid bulk modulus. Projection, fractional-step, and other stabilized integration strategies in the literature provide relevant alternatives \citep{morikawa2022,lian2023,wang2026}, while changes to the present time integration would require renewed validation of both pressure and boundary response.

The slopes also combine stress diffusion and periodic pressure filtering, whereas both operations are disabled in the principal elastic benchmarks. Consequently, those benchmark errors do not establish the accuracy of the regularized slope formulation. The shared stress/density-diffusion amplitude and the Shepard interval should be varied in matched calculations. In particular, preserving a constant excess-pressure field through normalization does not guarantee global pore-water-mass conservation, and the prescribed gravity increment in stress diffusion is not a general saturated-equilibrium solution. These properties require dedicated assessment before pressure smoothing or altered failure kinematics can be attributed to physical drainage alone.

\section{Conclusions}
A single-layer $u$--$p_w$ pressure-rate formulation has been integrated into GeoDualSPHysics to connect skeleton deformation, Darcy flow, and pore-pressure feedback. The extension adds hydraulic ghost-pressure recovery and drained-surface treatment to the existing boundary infrastructure, with independent mechanical slip settings and hydraulic drainage choices. This implementation focus distinguishes the work from the introduction of the underlying PR equations.

The 1D and 2D calculations reproduce pressure generation and dissipation under their specified loading and drainage sequences, while exposing remaining settlement and late-time pressure errors. The 3D Cryer sweep captures the Mandel--Cryer pressure rise for four Poisson ratios, with normalized pressure RMSE between 0.012 and 0.030 and peak underprediction of approximately 1.8--2.6\%. The sensitive-clay applications illustrate sequential retrogression and large post-failure deformation in 5 m and 8 m geometries. Their present role is qualitative validation against established failure mechanisms rather than a completed quantitative assessment of hydraulic influence.

Further work should prioritize joint pressure-and-settlement convergence, long-time hydraulic boundary accuracy, quantitative characterization of slope kinematics, and comparisons between matched coupled and undrained slope models. These checks will establish the extension's quantitative reliability across large-deformation problems.

\paragraph{Use of generative AI.}
OpenAI Codex was used to assist with manuscript drafting and language editing. The authors take responsibility for the scientific content and final text.

\paragraph{Data and code availability.}
The base GeoDualSPHysics solver is described by \citet{feng2026geodual}; its published software record is available at \url{https://doi.org/10.17632/z4sh62y97g.1}.

\FloatBarrier
\clearpage
\bibliographystyle{unsrtnat}
{\small\setlength{\bibsep}{2pt}\bibliography{references}}

\begin{thebibliography}{20}
\providecommand{\natexlab}[1]{#1}
\providecommand{\url}[1]{\texttt{#1}}
\expandafter\ifx\csname urlstyle\endcsname\relax
  \providecommand{\doi}[1]{doi: #1}\else
  \providecommand{\doi}{doi: \begingroup \urlstyle{rm}\Url}\fi

\bibitem[Jin and Yin(2022)]{jin2022}
Yin-Fu Jin and Zhen-Yu Yin.
\newblock Two-phase {PFEM} with stable nodal integration for large deformation
  hydromechanical coupled geotechnical problems.
\newblock \emph{Computer Methods in Applied Mechanics and Engineering},
  392:\penalty0 114660, 2022.
\newblock \doi{10.1016/j.cma.2022.114660}.

\bibitem[Bui et~al.(2008)Bui, Fukagawa, Sako, and Ohno]{bui2008}
Ha~H. Bui, Ryoichi Fukagawa, Kazunari Sako, and Shintaro Ohno.
\newblock Lagrangian meshfree particles method ({SPH}) for large deformation
  and failure flows of geomaterial using elastic--plastic soil constitutive
  model.
\newblock \emph{International Journal for Numerical and Analytical Methods in
  Geomechanics}, 32\penalty0 (12):\penalty0 1537--1570, 2008.
\newblock \doi{10.1002/nag.688}.

\bibitem[Blanc and Pastor(2012)]{blanc2012}
T.~Blanc and M.~Pastor.
\newblock A stabilized fractional step, {Runge--Kutta Taylor SPH} algorithm for
  coupled problems in geomechanics.
\newblock \emph{Computer Methods in Applied Mechanics and Engineering},
  221--222:\penalty0 41--53, 2012.
\newblock \doi{10.1016/j.cma.2012.02.006}.

\bibitem[Bui and Fukagawa(2013)]{bui2013}
Ha~H. Bui and Ryoichi Fukagawa.
\newblock An improved {SPH} method for saturated soils and its application to
  investigate the mechanisms of embankment failure: Case of hydrostatic
  pore-water pressure.
\newblock \emph{International Journal for Numerical and Analytical Methods in
  Geomechanics}, 37\penalty0 (1):\penalty0 31--50, 2013.
\newblock \doi{10.1002/nag.1084}.

\bibitem[Bui and Nguyen(2017)]{bui2017}
Ha~H. Bui and Giang~D. Nguyen.
\newblock A coupled fluid-solid {SPH} approach to modelling flow through
  deformable porous media.
\newblock \emph{International Journal of Solids and Structures}, 125:\penalty0
  244--264, 2017.
\newblock \doi{10.1016/j.ijsolstr.2017.06.022}.

\bibitem[Morikawa and Asai(2022)]{morikawa2022}
Daniel~S. Morikawa and Mitsuteru Asai.
\newblock Soil-water strong coupled {ISPH} based on $u$--$w$--$p$ formulation
  for large deformation problems.
\newblock \emph{Computers and Geotechnics}, 142:\penalty0 104570, 2022.
\newblock \doi{10.1016/j.compgeo.2021.104570}.

\bibitem[Feng et~al.(2022)Feng, Fourtakas, Rogers, and
  Lombardi]{feng2022coupled}
Ruofeng Feng, Georgios Fourtakas, Benedict~D. Rogers, and Domenico Lombardi.
\newblock Two-phase fully-coupled smoothed particle hydrodynamics ({SPH}) model
  for unsaturated soils and its application to rainfall-induced slope collapse.
\newblock \emph{Computers and Geotechnics}, 151:\penalty0 104964, 2022.
\newblock \doi{10.1016/j.compgeo.2022.104964}.

\bibitem[Lian et~al.(2023)Lian, Bui, Nguyen, and Haque]{lian2023}
Yanjian Lian, Ha~H. Bui, Giang~D. Nguyen, and Asadul Haque.
\newblock An effective and stabilised $(u-p_l)$ {SPH} framework for large
  deformation and failure analysis of saturated porous media.
\newblock \emph{Computer Methods in Applied Mechanics and Engineering},
  408:\penalty0 115967, 2023.
\newblock \doi{10.1016/j.cma.2023.115967}.

\bibitem[Chen et~al.(2024)Chen, Leung, Mori, Uchida, and Takumi]{chen2024}
Xiaoyu Chen, Yat~Fai Leung, Hirotoshi Mori, Shun Uchida, and Kazuhiro Takumi.
\newblock Single-layer soil-water coupled {SPH} method and its application to
  sinkhole simulation.
\newblock \emph{Acta Geotechnica}, 19:\penalty0 991--1018, 2024.
\newblock \doi{10.1007/s11440-023-02063-4}.

\bibitem[Feng et~al.(2024)Feng, Fourtakas, Rogers, and
  Lombardi]{feng2024coupled}
Ruofeng Feng, Georgios Fourtakas, Benedict~D. Rogers, and Domenico Lombardi.
\newblock A general smoothed particle hydrodynamics ({SPH}) formulation for
  coupled liquid flow and solid deformation in porous media.
\newblock \emph{Computer Methods in Applied Mechanics and Engineering},
  419:\penalty0 116581, 2024.
\newblock \doi{10.1016/j.cma.2023.116581}.

\bibitem[Yao et~al.(2025)Yao, Fourtakas, Rogers, and Lombardi]{yao2025}
C.~Yao, G.~Fourtakas, B.~D. Rogers, and D.~Lombardi.
\newblock 2-{D} {SPH} modelling of poroelasticity: $u$--$w$--$p$ and $u$--$p$
  formulations with absorbing boundary conditions and volumetric locking
  treatments.
\newblock \emph{Computers and Geotechnics}, 179:\penalty0 107016, 2025.
\newblock \doi{10.1016/j.compgeo.2024.107016}.

\bibitem[Wang and Wong(2026)]{wang2026}
Zhaonan Wang and Louis Ngai~Yuen Wong.
\newblock Unified drag-force temporal discretization in the $u$--$u$--$p$
  fractional step method.
\newblock \emph{Computers and Geotechnics}, 192:\penalty0 107910, 2026.
\newblock \doi{10.1016/j.compgeo.2026.107910}.

\bibitem[Tong et~al.(2026)Tong, Gu, Liu, Xia, and Zhang]{tong2026}
Tiancheng Tong, Xin Gu, Panyong Liu, Xiaozhou Xia, and Qing Zhang.
\newblock A stabilized and first-order consistent smoothed particle
  hydrodynamics for coupled flow-deformation analysis of saturated porous
  media.
\newblock \emph{International Journal for Numerical and Analytical Methods in
  Geomechanics}, 50:\penalty0 3739--3753, 2026.
\newblock \doi{10.1002/nag.70300}.

\bibitem[del Castillo et~al.(2026)del Castillo, Borja, and
  F{\'a}vero~Neto]{castillo2026}
Enrique~M. del Castillo, Ronaldo~I. Borja, and Alomir~H. F{\'a}vero~Neto.
\newblock A coupled $u$--$p_w$ {SPH} formulation for hydromechanical modeling
  of retrogressive landslides and comparison with a penalty-based approach.
\newblock \emph{International Journal for Numerical and Analytical Methods in
  Geomechanics}, 2026.
\newblock \doi{10.1002/nag.70322}.
\newblock Advance online publication.

\bibitem[Feng et~al.(2026)Feng, Zhao, Fourtakas, and Rogers]{feng2026geodual}
Ruofeng Feng, Jidong Zhao, Georgios Fourtakas, and Benedict~D. Rogers.
\newblock {GeoDualSPHysics}: a high-performance {SPH} solver for large
  deformation modelling of geomaterials with two-way coupling to multi-body
  systems.
\newblock \emph{Computer Physics Communications}, 320:\penalty0 109965, 2026.
\newblock \doi{10.1016/j.cpc.2025.109965}.

\bibitem[English et~al.(2022)English, Dom{\'i}nguez, Vacondio, Crespo, Stansby,
  Lind, Chiapponi, and G{\'o}mez-Gesteira]{english2022}
A.~English, J.~M. Dom{\'i}nguez, R.~Vacondio, A.~J.~C. Crespo, P.~K. Stansby,
  S.~J. Lind, L.~Chiapponi, and M.~G{\'o}mez-Gesteira.
\newblock Modified dynamic boundary conditions ({mDBC}) for general-purpose
  smoothed particle hydrodynamics ({SPH}): application to tank sloshing, dam
  break and fish pass problems.
\newblock \emph{Computational Particle Mechanics}, 9:\penalty0 911--925, 2022.
\newblock \doi{10.1007/s40571-021-00403-3}.

\bibitem[Feng et~al.(2021)Feng, Fourtakas, Rogers, and
  Lombardi]{feng2021stress}
Ruofeng Feng, Georgios Fourtakas, Benedict~D. Rogers, and Domenico Lombardi.
\newblock Large deformation analysis of granular materials with stabilized and
  noise-free stress treatment in smoothed particle hydrodynamics ({SPH}).
\newblock \emph{Computers and Geotechnics}, 138:\penalty0 104356, 2021.
\newblock \doi{10.1016/j.compgeo.2021.104356}.

\bibitem[Zhao et~al.(2019)Zhao, Bui, Lemiale, Nguyen, and Darve]{zhao2019}
Shaohan Zhao, Ha~H. Bui, Vincent Lemiale, Giang~D. Nguyen, and Felix Darve.
\newblock A generic approach to modelling flexible confined boundary conditions
  in {SPH} and its application.
\newblock \emph{International Journal for Numerical and Analytical Methods in
  Geomechanics}, 43:\penalty0 1005--1031, 2019.
\newblock \doi{10.1002/nag.2918}.

\bibitem[Terzaghi(1943)]{terzaghi1943}
Karl Terzaghi.
\newblock \emph{Theoretical Soil Mechanics}.
\newblock John Wiley \& Sons, New York, 1943.

\bibitem[Cryer(1963)]{cryer1963}
C.~W. Cryer.
\newblock A comparison of the three-dimensional consolidation theories of
  {Biot} and {Terzaghi}.
\newblock \emph{The Quarterly Journal of Mechanics and Applied Mathematics},
  16\penalty0 (4):\penalty0 401--412, 1963.
\newblock \doi{10.1093/qjmam/16.4.401}.

\end{thebibliography}
\end{document}